\documentclass[letterpaper]{article}
\usepackage{iccc}

\usepackage{hyperref}
\usepackage{url}
\usepackage[pdftex]{graphicx}
\usepackage{subcaption}
\usepackage{bm}
\usepackage{multirow,booktabs}
\usepackage{amsmath,amsfonts,bm}
\usepackage{comment}

\usepackage{times}
\usepackage{helvet}
\usepackage{courier}
\title{Score and Lyrics-Free Singing Voice Generation}

\author{Jen-Yu Liu$^1$, Yu-Hua Chen$^{1,2}$, Yin-Cheng Yeh$^1$ \& Yi-Hsuan Yang$^{1,2}$ \\
$^1$Taiwan AI Labs, Taipei, Taiwan \\
$^2$Academia Sinica, Taipei, Taiwan \\
 \\
\texttt{jyliu@ailabs.tw, cloud60138@citi.sinica.edu.tw, yyeh@ailabs.tw, yang@citi.sinica.edu.tw} \\
}
\begin{document} 
\maketitle
\begin{abstract}
Generative models for singing voice have been mostly concerned with the task of ``singing voice synthesis,'' i.e., to produce singing voice waveforms given musical scores and text lyrics. In this work, we explore a novel yet challenging alternative: singing voice generation without pre-assigned scores and lyrics, in both training and inference time. In particular, we outline three such generation schemes, and propose a pipeline to tackle these new tasks. Moreover, we implement such models using generative adversarial networks and evaluate them both objectively and subjectively.
\end{abstract}

\section{Introduction} 
\label{sec:intro}

The task of computationally producing singing voices is usually referred to as singing voice \emph{synthesis} (SVS) in the literature \cite{cook96cmj}.
Most researchers assume that the note sequence and the lyrics of the audio to be generated are given as the model input,
and aim to build synthesis engines that sound as natural and expressive as a real singer (e.g.,  \cite{hono19,lee19interspeech}).
As such, the content of the produced singing voice is largely determined by the given model input, which is usually assigned by human. 

However, singing according to a pre-assigned musical score and lyrics is only a part of the human singing activities. 
For example, we learn to spontaneously sing when we were children \cite{dowling84}. We do not need a score to sing when we are humming on the road or while taking a shower. The voices sung do not have to be intelligible. Jazz vocalists can improvise according to a chord progression, an accompaniment, or even nothing. 

We aim to explore such a new task in this paper: 
teaching a machine to sing with a training collection of singing voices, but without the corresponding musical scores and lyrics of the training data. Moreover, the machine has to sing without pre-assigned score and lyrics as well even in the inference (generation) time.  
This task is challenging in that, as the machine sees no lyrics at all, it hardly has any knowledge of the human language to pronounce or articulate either voiced or unvoiced sounds.
And, as the machine sees no musical scores at all, it has to find its own way learning the language of music in creating plausible vocal melodies.

Specifically, we consider three types of such score- and lyrics-free singing 
\emph{generation} tasks.
A \emph{free singer} sings with only random noises as the input. An \emph{accompanied singer} learns to sing over a piece of instrumental music, which is given as an audio waveform (again without score information). Finally, a \emph{solo singer} also sings with only noises as the input, but it uses the noises to firstly generate some kind of `inner ideas' of what to sing.
From a technical point of view, we can consider SVS as a \emph{strongly conditioned} task for generating singing voices, as the target output is  well specified by the input. In contrast, the proposed tasks are either 
\emph{unconditioned} or \emph{weakly conditioned}.
While our models are presumably more difficult to train than SVS models, they enjoy more freedom in the generation output, which may be desirable considering the artistic nature of singing.

The proposed tasks are challenging in a few aspects. 
\begin{itemize}
    \item 
    First, the tasks are 
    unsupervised as we do not provide any labels (e.g., labels of phonemes or pitches) 
    for the training singing files. The machine has to learn the complex structure of music directly from the audio signals. 
    \item Second, for training the free singer, unaccompanied vocal tracks are needed. As for the accompanied singer, we need additionally an accompaniment track for each vocal track. However, it is hard to amass such multi-track music data from the public domain. 
    \item Third,  for the accompanied singer case, there is no single ``ground truth'' and the relationship between the model input 
    and output 
    may be one-to-many. 
    This is because there are plenty of valid ways to sing over an accompaniment track.
    For diversity and artistic freedom, we cannot ask the machine to generate any specific singing voice in response to an accompaniment track, even if we have paired data of vocal and accompaniment tracks. 
\end{itemize}

To address the first and third issues, we explore the use of generative adversarial network (GAN), 
in particular conditional GAN \cite{cgan} to retain the possibility of generating singing voices with multiple modes.
Specifically, we design a novel GAN-based architecture to learn to generate the mel-spectrogram of singing voice, and then use 
a vocoder 
to generate the audio waveform. Rather than considering the mel-spectrograms as a fixed-size image,
we use gated recurrent units (GRUs) 
and dilated convolutions 
in both the generator and discriminator, to model both the local and sequential patterns in music and to facilitate the generation of variable-length waveforms.

\begin{figure}
	\centering
	\includegraphics[width=\columnwidth]{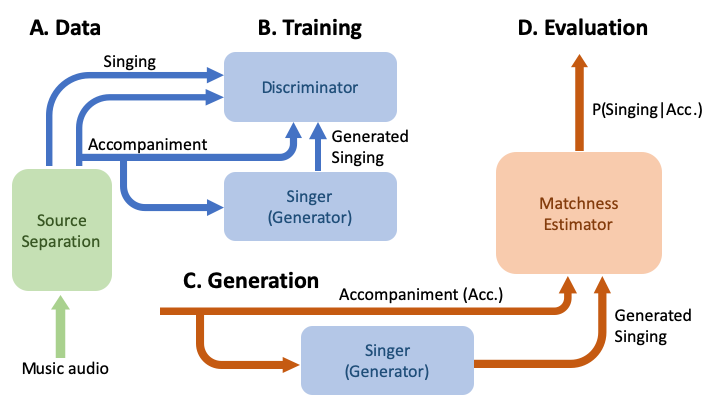}
	\caption{A pipeline for building the accompanied singer. We use source separation to get separated singing voice and accompaniment from
	professionally recorded audio files. Then, we use the separated tracks to train the generators and discriminators of GAN. For inference, we feed an unseen accompaniment to the trained singer model and let it ``sing.'' }
	\label{fig:overview}
\end{figure}

To address the second issue, we choose to implement a vocal source separation model with state-of-the-art separation quality \cite{liu19} for data preparation. 
The advantage of having a vocal separation model is that we can use as many audio files as we have to compile the training data. 
The proposed pipeline for training 
an accompanied singer is illustrated in Figure \ref{fig:overview}.



We implement the proposed singing voice generation models with a collection of Jazz music. For evaluation, we employ a few objective metrics and conduct a user study.
Samples of the generated singing voices can be found 
at \url{https://bit.ly/2mIvoIc}.
To demonstrate the use case of using the generated sounds as a sound source, we manually make a song in the style of Jazz Hiphop by sampling the output of a free singer we trained. This song can be heard at 
\url{https://bit.ly/2QkUJoJ}. For reproducibility, we release our code at \url{https://github.com/ciaua/score_lyrics_free_svg}.  

\section{Schemes of singing voice generation}
\label{sec:singers}

A \textbf{free singer} takes no conditions at all as the input. We want it to sing freely. The singing voices from a free singer may not even sound good, but they should sound like singing voice. A free singer is like we are freely humming or singing on the road walking or in the bathroom taking a shower. We may not even know what we are singing and likely there is no underlying musical score. 


An \textbf{accompanied singer} takes as the input a sequence of accompaniment-derived features. An accompanied singer tries to generate singing voices that match the accompaniment track in some way. It is similar to the case of Karaoke, 
where a backing accompaniment track is played from a speaker, the lyrics and a video are displayed on a screen, and a user tries to sing according to the lyrics and the backing track. The difference is that, this time the user is a trained model and we do not ask it to follow the lyrics or the exact pitch contour of the accompaniment. 
The note sequence found in the singing has to be in harmony with, but not a duplicate of, that in the backing track.


A \textbf{solo singer} is similar to a free singer in that both takes no conditions as the input. However, a solo singer would generate an `inner idea' first, and then sing according to that. 
The inner idea 
can take several forms. In this work, we instantiate this scheme with the inner idea 
being a chord progression (namely a sequence of chord labels). The distribution of inner ideas 
is modeled by an auto-regressive recurrent network we build for chord progression generation. 

Alternatively, we can think of a solo singer as a combination of an idea generator and an an accompanied singer.
For an accompanied singer, the information extracted from the given accompaniment track can take several forms such as transcribed pitches and chord progressions. 
A solo singer learns to generate such information on its own, without reference to an actual accompaniment track.

\section{Models}
\label{sec:models}

To account for the absence of supervised data and the highly complicated spatio-temporal patterns in audio spectrograms, we propose a new adversarial net that features heavy use of 
GRUs \cite{cho14}, dilated convolutions \cite{oord16}, 
and feature grouping to build our singer models. We provide the algorithmic details below. 

\subsection{Block of GRU-Grouped Dilated Convolution-Group Normalization}
\label{sec:g3block} 
Network architectures with stacked blocks of GRUs and dilated convolutions have been used to attain state-of-the-art performance in blind musical source separation \cite{liu19}. In a source separation task, a model learns to decompose, or unmix, different sources (e.g., vocal, piano, bass, drum) from a mixture signal. 
This requires the abilities to model the relationships between different sources as well as the relationships between neighboring time frames. 
The output spectrograms are also expected to be distortion-less and of high audio quality. 
For it has demonstrated its capability in source separation, we adopt it as a building block of the singer models. Especially, we want the singer models to also consider accompaniment information. 

Specifically, one such block we adopted in our models is a stack of GRU, dilated convolution with feature grouping, and group normalization \cite{wu18}. The input to the GRU, the output of the GRU, and the output of the group normalization are summed to form the output of the block. We note that the original `D2 block' used in \cite{liu19} uses dilated GRU and uses weight normalization \cite{salimans16} for the dilated convolution layers. However, empirically we find that it is easier for the singer models to converge by replacing weight normalization with group normalization, and using plain GRUs is as good as using dilated GRUs.   We refer to our 
blocks as GRU-grouped dilated convolution-group normalization block (`G3 block').

\subsection{Singer Models with BEGAN, G3 blocks and Frame-wise Noises (G3BEGAN)}
\label{sec:began}

The accompanied singers and solo singers have to take conditions as part of their input. One desirable property of the models is the ability to generate voices with arbitrary length, as the conditional signal can be of variable length. 
Besides, the model has to deal with the one-to-many issue mentioned in the introduction, and the absence of supervisory signals. With these in mind, we
design a GAN architecture for score and lyrics-free voice generation.
In particular, we pay special attention to the following three components: 1) the network architecture, 2) the input noises for GAN, and 3) the loss function of the discriminator.

Let us first take a look at two existing GAN  models for audio generation: \cite{gansynth} and \cite{donahue19iclr}. Their generators and discriminators are both based on 2D convolutions, transposed 2D convolutions and dense (linear) layers. The generators take a vector $\textbf{z} \in \mathbb{R}^U$ as the input noise and use transposed convolutions to expand $\textbf{z}$ so that a temporal dimension emerges in the expanded intermediate matrices. The number of temporal frames in the final output depends on the total strides used in all the transposed convolutions. The discriminators take the output of the generators or the real signal as the input, and compress the input matrix with convolution layers until the output becomes a single value represents the prediction of true (real) or false (generated) data. 

A main reason why
existing models cannot generate variable-length output is the need to expand $\mathbf{z}$ by transposed convolution layers.
We remedy this by using an architecture consisting of the proposed G3 blocks, and convolutions without strides, for both the generators $G(\cdot)$ and discriminators $D(\cdot)$.  
Moreover, instead of using a single noise vector, our models take as input a sequence of noise vectors, denoted as 
$\textbf{Z} \in \mathbb{R}^{U \times T}$, that has the same temporal length as the desired output $\textbf{Y}$.
Each column of $\textbf{Z}$ is sampled independently from a Gaussian distribution $Normal(0, 1)$.
At the first glance, it might feel unnatural to have one noise vector per frame as that may result in fast oscillations in the noises. However, we note that the output of $G(\cdot)$ for the $t$-th frame depends not only on the $t$-th column of $\mathbf{Z}$ (and $\mathbf{C}$ or $\mathbf{I}$), but the entire $\mathbf{Z}$ (and the condition matrices),
due to the recurrent GRUs in the model. We expect that the GRUs in the discriminator $D(\cdot)$ would force $G(\cdot)$ to generate consistent consecutive frames. 
Therefore, the effect of the frame-wise noises might be introducing variations to the generation result, for example by adjusting the modes of the generated frame-wise features.

As for the loss function of $D(\cdot)$, 
we experiment with the following three options: the vanilla GAN,
the LSGAN \cite{lsgan} that adopts the least squares loss function for the discriminator, and the boundary equilibrium GAN (BEGAN) \cite{berthelot17} that adopts an ``auto-encoder style'' discriminator loss.
 The $D(\cdot)$ in either GAN or LSGAN is implemented as a classifier aiming to distinguish between real and generated samples, whereas the $D(\cdot)$ in BEGAN is an autoencoder aiming to reconstruct its input. 
Specifically, in BEGAN, the loss functions $l_D$ and $l_G$ for the discriminator and generator, as in the case of the accompanied singer, are respectively:
\begin{align}
    l_D=& L(\mathbf{X}, \mathbf{C})-\tau_s L(G(\mathbf{Z}, \mathbf{C}), \mathbf{C}) \,, \label{eq:1}\\ 
    l_G=& L(G(\mathbf{Z}, \mathbf{C}), \mathbf{C}) \,, \label{eq:2} 
\end{align}
where $\mathbf{X} \in \mathbb{R}^{K \times T}$ is the feature sequence of a real vocal track sampled from the training data, $G(\mathbf{Z},\mathbf{C})  \in \mathbb{R}^{K \times T}$ is the feature sequence for the generated vocal track, and $L(\cdot)$ is a function that measures how well the discriminator $D(\cdot)$, implemented as an auto-encoder, reconstructs its input:
\begin{equation}
L(\mathbf{M}, \mathbf{C})=\frac{1}{WT}\sum_{w,t}|D(\mathbf{M}, \mathbf{C})_{w, t}-M_{w,t}|, 
\end{equation}
for an arbitrary $W\times T$ matrix $\mathbf{M}$,
where we use $M_{w,t}$ to denote the $(w, t)$-th element of a matrix $\mathbf{M}$ (and similarly for $D(\mathbf{M}, \mathbf{C})_{w, t}$).
Moreover, the variable $\tau_s$ in Eq. (\ref{eq:1}) is introduced by BEGAN to balance the power of $D(\cdot)$ and $G(\cdot)$ during the learning process. 
It is dynamically set to be
$\tau_{s+1}=\tau_s+\lambda(\gamma L(\mathbf{X}, \mathbf{C})-L(G(\mathbf{Z}, \mathbf{C}), \mathbf{C}))$, for each training step $s$, with $\tau_s \in [0,1]$. $\lambda$ and $\gamma$ are manually-set hyperparameters.

Our pilot study (not reported here due to space restriction) shows that BEGAN performs the best. Therefore, we consider below the BEGAN-based model, dubbed G3BEGAN.
See Table \ref{tab:g3began} for some details of the network architecture.

\begin{table}
\centering
\begin{tabular}{c|l} 
\toprule
& {\bf Details} \\ 
\toprule
\multirow{3}{*}{\bf Input}       & 1DConv (k$=$3, d$=$1) \\ 
            & Group normalization (g$=$4) \\
            & Leaky ReLU (0.01)\\
\midrule
\multirow{4}{*}{\bf G3 Blocks 1}  & GRU \\ 
            & Grouped 1DConv (k$=$3, d$=$2, g$=$4) \\ 
            & Group normalization (g$=$4) \\
            & Leaky ReLU (0.01) \\
\midrule
\multirow{4}{*}{\bf G3 Block 2}  & GRU    \\ 
            & Grouped 1DConv (k$=$3, d$=$2, g$=$4) \\ 
            & Group normalization (g$=$4) \\
            & Leaky ReLU (0.01) \\
\midrule
{\bf Output}      & 1DConv (k$=$3, d$=$1) \\ 
\bottomrule
\end{tabular}
\caption{Network architecture of the generator and the discriminator of the proposed G3BEGAN model, each uses two G3 blocks. We use `k', `d', and `g' to denote the kernel size, dilation length, and number of groups.
}
\label{tab:g3began}
\end{table}




\subsection{Source Separation (SS)}

To get the vocal tracks for training our singer models,
we implement a source separation (SS) model following the architecture proposed by \cite{liu19}, which represents the state-of-the-art as evaluated on the MUSDB dataset \cite{musdb18}. 
MUSDB contains clean vocal and accompaniment tracks for various genres.
As we are mainly interested in Jazz music in this work, 
we collect additionally 4.5 hours of 
Jazz piano solo audio to augment  the MUSDB for training our SS model.
Our model can isolate out not only the vocal track but also the piano track from an arbitrary song.
Please visit \url{https://bit.ly/2Xattua} for samples of the separation result of our SS model.

\subsection{Chord Generator}
\label{appendix:chord_generator}
For training the solo singer, we implement a chord generator in this work. 
It is aimed to generate chord progressions freely under some given conditions. It supports 12 major and 12 minor keys, 10 tempo options from 60 to 240 BPM, 6 time signature options, and 51 chord qualities (612 chords in total).  The conditions, key, tempo, and time signatures, are encoded into one-hot representation and concatenated together as a 40-dimension vector.  Our chord generator mainly consists with 3 stacked GRU layers. 
The input of each time step is a 524-dimensional vector consisting of a chord embedding and a beat-related one-hot positional encoding, to encourage the model to follow certain rhythmical pattern.  This input array passes through a fully-connected layer to 512-dimension and is used as the input of the GRUs.  The training data are the lead sheets from the Wikifonia dataset.  We augmented the data by rotating the keys, leading to in total 80,040 lead sheets for training.

\section{Experiments}
\label{sec:exp}

\subsection{Implementation Details} 
\label{sec:exp:setup}

We use 80-dimensional mel-spectrograms as the acoustic features modeled and generated by the singer models. 
We use the python package \texttt{librosa} \cite{mcfee2015librosa}, with default settings, to compute the mel-spectrograms from audio.
A mel-spectrogram is passed to a WaveRNN vocoder \cite{kalchbrenner18} to generate an audio signal from mel-spectrograms. 
Our implementation of the WaveRNN vocoder is based on the code from Fatchord.\footnote{\url{https://github.com/fatchord/WaveRNN}} Instead of using off-the-shelf pre-trained vocoders, which are typically trained for text-to-speech  (TTS), we train our vocoder from scratch with a set of 3,500 vocal tracks separated by our SS model from an in-house collection of music that covers diverse musical genres.

We collect 17.4 hours of Jazz songs containing female voices and 7.6 hours of Jazz songs with male voices.
We again use our SS  model to get the vocal tracks from these songs.
For batched training, we divide the tracks into 10-second sub-clips. Sub-clips that contain less than 40\% vocals, as measured from energy, are removed.
This leads to 9.9-hour and 5.0-hour training data for female and male Jazz vocals, respectively. 
200 and 100 sub-clips are reserved from the training set as the validation set for female singing and male singing, respectively. 
Singer models with G3BEGAN are trained with Adam \cite{adam} with $10^{-4}$ learning rate, mini-batch size 5, and gradient norm clipping with magnitude 3. We train the model for 500 epochs, and then pick  
the epoch
with the best convergence rate \cite{berthelot17} for evaluation.

For the accompanied singer, we experiment with extracting pitch-related information from the accompaniment track 
to condition 
the generation of the vocal track. The assumption here is that whether the generated vocal track is in harmony with the accompaniment track can be largely determined by pitch-related information. 
For this purpose, we implement a piano transcription model to transcribe the separated piano track, leading to 88-dimensional transcribed frame-wise pitch as the accompaniment condition.
We implement our piano transcription model with the G3 blocks, following the training procedure of \cite{hawthorne18ismir}. 
We note that, the clips in the training set of our singer models may not contain piano playing. 
Even if a clip contains piano playing, the piano may not play across the entire clip. Hence, the  models have to learn to sing either with or without the piano accompaniment.

For evaluating the accompanied singer, we collect 5.3 hours of Jazz music from Jamendo.\footnote{\url{https://www.jamendo.com}} As the music hosted on Jamendo are copyright-free, we will later be able to share the test set to the research community.
We apply our SS model to the audios to get the piano track,  divide each track into 20-second sub-clips,\footnote{Please note that this is longer than the 10-second sub-clips we used to train the singer models. This is okay as our model can generate variable-length output.} and discard those clips that are silent (i.e., do not contain piano). Piano transcription is also applied to the separated piano track, yielding 402 20-second sub-clips for evaluation. 
As for evaluating the solo singer,
we generate 402 chord progressions by our chord generator. 

\begin{table*}[t]
\begin{center}
\begin{tabular}{l|cccc}
\toprule
\multirow{2}{*}{{\bf Proposed model}}  & \multicolumn{2}{c}{\bf Average pitch (Hz)} & {\bf Vocalness} & \multirow{2}{*}{{\bf Matchness}} \\ 
\cline{2-4}
               & {\bf CREPE}  & {\bf JDC} & {\bf JDC} &  \\ 
\midrule
Free singer (female)        & 288 $\pm$ 28 & 292 $\pm$ 28 & 0.48 $\pm$ 0.09 & --13.28 $\pm$ 3.80~~ \\
Accompanied singer (female) & 313 $\pm$ 18 & 316 $\pm$ 19 & 0.55 $\pm$ 0.11 & --9.25 $\pm$ 3.13 \\ 
Solo singer (female)        & 302 $\pm$ 17 & 306 $\pm$ 18 & 0.56 $\pm$ 0.10 & --9.30 $\pm$ 3.11 \\
\midrule
Free singer (male)          & 248 $\pm$ 39 & 242 $\pm$ 32 & 0.44 $\pm$ 0.16 & --13.29 $\pm$ 3.19~~  \\
Accompanied singer (male)   & 207 $\pm$ 14 & 200 $\pm$ 15 & 0.44 $\pm$ 0.13 & --9.31 $\pm$ 3.16 \\ %
Solo singer (male)          & 213 $\pm$ 14 & 207 $\pm$ 16 & 0.46 $\pm$ 0.12 & --9.30 $\pm$ 3.13 \\
\toprule
\bf{Baseline: Singing voice synthesis} \\
\midrule
Sinsy (training vocal, female)  & 305 $\pm$ 59 & 308 $\pm$ 57 & 0.71 $\pm$ 0.17 & --9.20 $\pm$ 3.12\\
Sinsy (training vocal, male)    & 260 $\pm$ 86 & 259 $\pm$ 72 & 0.73 $\pm$ 0.14 & --9.09 $\pm$ 3.14\\
Sinsy (testing piano skyline, female)  & ~~523 $\pm$ 138 & 431 $\pm$ 62 & 0.66 $\pm$ 0.14 & --8.88 $\pm$ 3.04 \\
Sinsy (testing piano skyline, male)    & ~~520 $\pm$ 137 & 423 $\pm$ 61 & 0.62 $\pm$ 0.15 & --8.93 $\pm$ 3.02 \\
\toprule
\bf{Baseline: Training data} \\
\midrule
Wikifonia: real melody-chords      & --- & --- & --- & --7.04 $\pm$ 2.91 \\
Wikifonia: random melody-chords & --- & --- & --- & --13.16 $\pm$ 3.72~~ \\
Singer train data (vocals, female) & ~~312 $\pm$ 70 & 310 $\pm$ 56 & 0.60 $\pm$ 0.14  & --9.24 $\pm$ 3.09 \\
Singer train data (vocals, male) & ~~263 $\pm$ 93 & 258 $\pm$ 75 & 0.64 $\pm$ 0.16 & --9.09 $\pm$ 3.22 \\
Singer train data (accomp., female) & --- & --- & 0.05 $\pm$ 0.09 & --- \\
Singer train data (accomp., male)   & --- & --- & 0.12 $\pm$ 0.15 & --- \\
MUSDB clean vocals &  ~~271 $\pm$ 81 & 283 $\pm$ 75 & 0.59 $\pm$ 0.14 & --- \\ 
\bottomrule
\end{tabular}
\end{center}
\caption{Result of objective evaluation for our singer models and a few baseline methods.}
\label{tab:singers_objective}
\end{table*}

\subsection{Baselines} \label{sec:baselines}

As this is a new task, there is no previous work that we can  compare with. Therefore, we establish the baselines by 1) computing the baseline objective metrics (see Section \ref{sec:metrics}) from the training data of the singing models, and 2) using existing SVS systems for synthesizing singing voices.

For the SVS baselines, we employ Sinsy \cite{ss_hmm2,sinsy18apsipa}
and Synthesizer V \cite{synthV},
the two well-known SVS systems that are publicly accessible.
For Sinsy, we use the publicly available repository\footnote{\url{https://github.com/mathigatti/midi2voice}} to query the Sinsy API;\footnote{\url{http://sinsy.jp/}} we use the HMM version \cite{ss_hmm2} instead of the deep learning version as the latter cannot generate male voices.
For Synthesizer V, we use their software.\footnote{\url{https://synthesizerv.com/}}
We use Sinsy for both objective and subjective tests, but Synthesizer V for subjective test only, for the latter does not provide a functionality to batch process a collection of MIDI files and lyrics.

SVS systems require lyrics and melody as the input. 
For the lyrics, we choose to use multiple `la,' the default lyrics for Synthesizer V.\footnote{As our models do not contain meaningful lyrics, to be fair the baselines should not contain meaningful lyrics either. 
We choose `la' because people do sometimes sing with `la' and it has no semantic meaning. 
An alternative way to get the lyrics is by randomly sampling a number of characters. However, randomly sampling a reasonable sequence of characters is not a trivial task.} 
For the melodies, we adopt two methods:
\begin{itemize}
   \item Vocal transcription from singer training data. We use CREPE to transcribe the separated vocals from the singer training data, and convert it to MIDI format. 
   \item Piano transcription from the Jamendo testing data. 
   The  transcription result often contains multiple notes at the same time. Hence, we further use the skyline algorithm \cite{Ozcan2005} 
   to the  transcription result to get a melody line comprising the highest notes.
\end{itemize}

\subsection{Objective Metrics and Objective Evaluation Result}
\label{sec:metrics}

The best way to evaluate the performance of the singer models 
is to listen to the generated results. Therefore, we encourage  readers 
to listen to the audio files in our demo website, mentioned in the end of the introduction section.
However, objective evaluation remains desirable, either for model development or for gaining insights into the generation result. 
We propose the following metrics for our tasks.
\begin{itemize}
    \item \textbf{Vocalness} measures whether an audio clip contains singing voices. There are different publicly available tools for detecting singing voices in an audio mixture (e.g., \cite{lee18}).
    We choose the JDC model
    \cite{kum19as} for it represents the state-of-the-art. In this model, the pitch contour is also predicted in addition to the vocal activation. If the pitch at a frame is outside a reasonable human pitch range (73--988 Hz defined by JDC), the pitch is set to 0 at that frame. We consider a frame as being vocal if it has a vocal activation $\ge 0.5$ AND has a pitch $>0$. Moreover, we define the vocalness of an audio clip as the proportion of its frames that are vocal.
    The tool is applied to the non-silent part of an audio\footnote{The non-silent frames are derived by using the \texttt{librosa} function `effects.\_signal\_to\_frame\_nonsilent.'} of the generated singing voices only, excluding the accompaniment part.
    
    \item \textbf{Average pitch}: We estimate the pitch (in Hz) for each frame with two pitch detection models: the state-of-the-art monophonic pitch tracker CREPE \cite{kim18},
    and JDC.
    The average pitch is computed by averaging the pitches across the frames with confidence higher than $0.5$ for CREPE, and across the frames that are estimated to be vocal for JDC.
    
    \item \textbf{Singing-accompaniment matchness}: 
    To objectively measure matchness, we build a melody harmonization recurrent network model (MH) by adapting our chord generator, using additionally the melody tracks found in the Wikifonia dataset.  Specifically, the MH model intends to generate a chord sequence given a melody sequence. Such a model can be learned by using the pairs of melody and chord tracks in Wikifonia. We add the chroma representation of the melody with window size of a quarter-note to the input vector. 
    Given a pair of melody and chord sequences, the MH model computes the likelihood of observing that chord sequence as the output when taking the melody sequence as the model input. We use the average of the log likelihood across time frames as the matchness score. As the MH model considers symbolic sequences, we use CREPE 
    to transcribe the generated voices, and Madmom \cite{DBLP16BockKSKW} to recognize the chord sequence from the accompaniment track. 
\end{itemize}


\begin{table*}[t]
\begin{center}
\begin{tabular}{l|cccc}
\toprule
\multicolumn{1}{l|}{\bf Model (epochs trained)} & 
 \multicolumn{1}{c}{\bf Sound quality} & \multicolumn{1}{c}{\bf Vocalness}  &
 \multicolumn{1}{c}{\bf Expression}  &
 \multicolumn{1}{c}{\bf Matchness} \\
\midrule
G3BEGAN (20 epochs)      &  1.59 $\pm$ 0.82 & 1.93 $\pm$ 0.99 & 1.98 $\pm$ 0.88 & 2.18 $\pm$ 1.08 \\
G3BEGAN (240 epochs)      & 2.24 $\pm$ 0.93 & 2.66 $\pm$ 1.01 & 2.60 $\pm$ 1.01 & 2.58 $\pm$ 1.05 \\
G3BEGAN (final)      & 2.38 $\pm$ 0.96 & 2.98 $\pm$ 1.02 & 2.85 $\pm$ 1.00 & 2.74 $\pm$ 1.04 \\
\bottomrule
\end{tabular}
\end{center}
\caption{Mean opinion scores (MOS) and standard deviations in four evaluation criteria collected from the first user study, for different versions of accompanied singer (female).
The scores are in 5-point Likert scale (1--5); the higher the better.}
\label{tab:user_study_score}
\end{table*}

\begin{table*}[t]
\begin{center}
\begin{tabular}{l|cccc}
\toprule
\multicolumn{1}{l|}{\bf Model (epochs trained)} & 
 \multicolumn{1}{c}{\bf Sound quality} & \multicolumn{1}{c}{\bf Vocalness}  &
 \multicolumn{1}{c}{\bf Expression}  &
 \multicolumn{1}{c}{\bf Matchness} \\
\midrule
G3BEGAN (final)      &  1.71 $\pm$ 0.70 & 2.39 $\pm$ 1.11 & 2.27 $\pm$ 1.06 & 2.34 $\pm$ 1.16 \\
Sinsy \cite{ss_hmm2}      & 3.19 $\pm$ 1.07 & 2.90 $\pm$ 1.01 & 2.40 $\pm$ 0.98 & 2.10 $\pm$ 0.90 \\
Synthesizer V \cite{synthV}      & 3.57 $\pm$ 1.07 & 3.30 $\pm$ 1.24 & 3.25 $\pm$ 1.10 & 3.35 $\pm$ 1.15 \\
\bottomrule
\end{tabular}
\end{center}
\caption{MOS from the second user study, comparing our model and two existing SVS systems.}
\label{tab:user_study_score2}
\end{table*}

Several observations can be made from the result shown in Table \ref{tab:singers_objective}.
In terms of the average pitch, we can see that the result of our model is fairly close to that of the singing voices in the training data. Moreover, the average pitch of the generated female voices is higher than that of the generated male voices as expected.
We can also see that the Sinsy singing voices  tend to have overly high pitches, when the melody line is derived from a piano playing  (denoted as `testing piano skyline.').

In terms of vocalness, our models score in general lower than Sinsy, and the singing voices in the training data. However, the difference is not that far. As a reference, we also compute the vocalness of the accompaniments in the training set (denoted as `accomp.') and it is indeed quite low.\footnote{We note that Sinsy even scores higher in vocalness than the training data. This may be due to the fact that real singing voices are recorded under different conditions and effects.}

As for matchness,  we show in Table \ref{tab:singers_objective} the score computed from the real melody-chords pairs of Wikifonia (--7.04) and that from random pairs of Wikifonia (--13.16). 
We can see that the accompanied singers
score higher than the random baseline and the free singer as expected.\footnote{The matchness scores of the free singers are computed by pairing them with the 402 test clips.} 
Moreover, the matchenss scores of the accompanied singers are close to that of the singer training data.

From visually inspecting the generated spectrograms and listening to the audio, the models seem to learn the characteristics of the singing melody contour (e.g., the F0 is not stable over time). Moreover, the female singer models learn better than the male counterparts, possibly because of the larger training set.

\subsection{User Study and Subjective Evaluation Result} 
\label{sec:user_study}

We conduct two online 
user studies to evaluate the accompanied singer, the female one. In the first user study, we compare the `final' model (with the number of epochs selected according to a validation set) against two early versions of the model trained with less epochs. 
In the second one, we compare the proposed accompanied singer with Sinsy and Synthesizer V.

In the first study, we recruit 39 participants to each rate the generated singing 
for three different accompaniment tracks (each 20 seconds), one accompaniment track per page. The subjects are informed the purpose of our research (i.e., score and lyrics-free singing voice generation) and the user study (to compare three computer models), and are asked to listen in a quiet environment with proper headphone volume. No post-processing (e.g., noise removal, EQ adjustment) is applied to the audio. The ordering of the result of the three models is randomized.

The process of the second study is similar to the first one, but it includes five different accompaniments (randomly chosen from those used in the first user study) and the respective generated/synthesized singing voices. 
The melodies used for synthesis are those from the piano skyline of the test data, so that our model can be compared with the synthesis methods with the same accompaniment. 
A separate set of 21 subjects participate in this study. The audio files used in this user study can be downloaded from \url{https://bit.ly/2qNrekv}.

Tables \ref{tab:user_study_score} and \ref{tab:user_study_score2} show the result of the two studies. 
We can see that the model indeed learns better with more epochs.
Among the four evaluation criteria, the Sound Quality is rated lower than the other three in both studies, suggesting room for improvement.

By comparing the proposed model with the two SVS systems, we see that Synthesizer V performs the best for all the evaluation criteria. Our model achieves better Matchness than Sinsy, and achieves a rating close to Sinsy in Expression.\footnote{We note that Sinsy and Synthesizer V have an unfair advantage on matchness because their singing voices are basically synthesized according to the melody lines of the accompaniment. From Table \ref{tab:user_study_score2}, we see that Synthesizer V does exhibit this advantage, while Sinsy does not. We observe that the Sinsy singing voices do not always align with the provided scores. The fact that Synthesizer V has higher audio quality seem to promote its score in the other criteria. The presence of the result of  Synthesizer V seems to also make the subjects in the second study rate the proposed model lower than the subjects do in the first study.}
In general, we consider the result as  promising considering that our models are trained from scratch with little knowledge of human language.

\section{Related work}
\label{sec:related}

While early work on SVS is mainly based on digital signal processing (DSP) techniques 
such as sampling concatenation 
\cite{cook96cmj,bonada07spm}, 
machine learning approaches offer greater flexibility and have been more widely studied in recent years.
Hidden Markov models (HMMs), in particular, have been shown to work well for the task \cite{ss_hmm1}.  
The Sinsy system, a baseline model in Section \ref{sec:exp}, is also based on HMMs \cite{ss_hmm2}.
\cite{nishimura16interspeech} report improved  naturalness by using deep neural nets instead of HMMs. 
Since then, many neural network models have been proposed.

The model presented by  \cite{nishimura16interspeech} uses simple fully-connected layers to map symbolic features extracted from the user-provided scores and lyrics, to a vector of acoustic features for synthesis. The input and output features are time-aligned frame-by-frame beforehand by well-trained HMMs. The input features consist of score-related features (e.g., the key of the current bar and the pitch of the current musical note), and lyrics-related ones (the current phoneme identify, the number of phonemes in the current syllable, and the duration of the current phoneme).
The output features consist of spectral and excitation parameters and their dynamic features  \cite{sinsy18apsipa}, which altogether can then be turned into audio with a DSP technique called the MLSA filter \cite{mlsa83icassp}.

The aforementioned model has been extended in many aspects. For instance, using convolutional layers and recurrent layers in replacement of the fully-connected layers for learning the mapping between input and output features has been respectively investigated by \cite{nakamura19arxiv} and \cite{kim18interspeech}.
Using neural vocoders such as the WaveNet \cite{oord16} instead of the MLSA filter has been shown to improve naturalness by \cite{nakamura19arxiv}.
Rather than using hand-crafted features for the input and output, \cite{lee19interspeech} train a model to predict the mel-spectrogram directly from time-aligned lyrics and pitch labels, and then use the Griffin-Lim algorithm \cite{griffin-lim} to synthesize the audio. 
Modern techniques such as adversarial loss and attention module have also been employed \cite{lee19interspeech}.
Synthesizer V \cite{synthV}, the other baseline model we employ in Section \ref{sec:exp}, is based on a hybrid structure that uses both deep learning and sample-based concatenation.\footnote{\url{https://synthv.fandom.com/wiki/File:Synthesizer_V_at_the_Forefront_of_Singing_Synth} (last accessed: Nov. 12, 2019)}

While exciting progress has been made to SVS, the case of score and lyrics-free singing voice generation, to our best knowledge, has not been tackled thus far.
Similar to \cite{lee19interspeech}, we do not use hand-crafted features and we train our model to predict the mel-spectrograms.


\section{Conclusion}
\label{sec:conclusion}

In this paper, we have introduced a novel task of singing voice generation that does not use musical scores and lyrics. Specifically, we proposed three singing schemes with different input conditions: free singer,  accompanied singer, and solo singer. We have also proposed a BEGAN based architecture that uses GRUs and grouped dilated convolutions to learn to generate singing voices in an adversarial way.  
For evaluating such models, we proposed several objective metrics and implemented a model to measure the compatibility between a given accompaniment track and the generated vocal track. The evaluation shows that the audio quality of the generated voices still leave much room for improvement, but in terms of humanness and emotion expression our models work fine. 

Score and lyrics-free singing voice generation is a new task, and this work represents only a first step tackling it. 
How such models can contribute to computational creativity remains to be studied.
From a technical point of view, there are also many interesting ideas to pursue. For example, we have chosen to extract only pitch-related information from the accompaniment track for building the accompanied singer,  but a more interesting way is to let the model learns to extract relevant information on its own. 
In the near future, we plan to
investigate advanced settings that allow for timbre and expression control, and experiment with other network architectures, such as 
coupling a fine-grained auto-regressive model with a multiscale generation procedure as done in MelNet \cite{melnet}, 
or using multiple discriminators that evaluate the generated audio based on multi-frequency random windows as done in GAN-TTS \cite{gantts}.

\bibliographystyle{iccc}
\bibliography{jy20iccc}

\begin{thebibliography}{}

\bibitem[\protect\citeauthoryear{Berthelot, Schumm, and
  Metz}{2017}]{berthelot17}
Berthelot, D.; Schumm, T.; and Metz, L.
\newblock 2017.
\newblock Began: Boundary equilibrium generative adversarial networks.
\newblock {\em arXiv preprint}.

\bibitem[\protect\citeauthoryear{Bi\'{n}kowski \bgroup et al.\egroup
  }{2019}]{gantts}
Bi\'{n}kowski, M.; Donahue, J.; Dieleman, S.; Clark, A.; Elsen, E.; Casagrande,
  N.; Cobo, L.~C.; and Simonyan, K.
\newblock 2019.
\newblock High fidelity speech synthesis with adversarial networks.
\newblock {\em arXiv preprint arXiv:1909.11646}.

\bibitem[\protect\citeauthoryear{B{\"{o}}ck \bgroup et al.\egroup
  }{2016}]{DBLP16BockKSKW}
B{\"{o}}ck, S.; Korzeniowski, F.; Schl{\"{u}}ter, J.; Krebs, F.; and Widmer, G.
\newblock 2016.
\newblock madmom: a new python audio and music signal processing library.
\newblock {\em arXiv preprint arXiv:1605.07008}.

\bibitem[\protect\citeauthoryear{Bonada and Serra}{2007}]{bonada07spm}
Bonada, J., and Serra, X.
\newblock 2007.
\newblock Synthesis of the singing voice by performance sampling and spectral
  models.
\newblock {\em IEEE Signal Processing Magazine} 24(2):67--79.

\bibitem[\protect\citeauthoryear{Cho \bgroup et al.\egroup }{2014}]{cho14}
Cho, K.; van Merrienboer, B.; Bahdanau, D.; and Bengio, Y.
\newblock 2014.
\newblock On the properties of neural machine translation: Encoder-decoder
  approaches.
\newblock In {\em Proc. Workshop on Syntax, Semantics and Structure in
  Statistical Translation}.

\bibitem[\protect\citeauthoryear{Cook}{1996}]{cook96cmj}
Cook, P.~R.
\newblock 1996.
\newblock Singing voice synthesis: History, current work, and future
  directions.
\newblock {\em Computer Music Journal} 20(3):38--46.

\bibitem[\protect\citeauthoryear{Donahue, McAuley, and
  Puckette}{2019}]{donahue19iclr}
Donahue, C.; McAuley, J.; and Puckette, M.
\newblock 2019.
\newblock {Adversarial audio synthesis}.
\newblock In {\em Proc. International Conference on Learning Representations}.

\bibitem[\protect\citeauthoryear{Dowling}{1984}]{dowling84}
Dowling, W.~J.
\newblock 1984.
\newblock {Development of musical schemata in children's spontaneous singing}.
\newblock {\em Advances in Psychology} 19:145--163.

\bibitem[\protect\citeauthoryear{Engel \bgroup et al.\egroup }{2019}]{gansynth}
Engel, J.; Agrawal, K.~K.; Chen, S.; Gulrajani, I.; Donahue, C.; and Roberts,
  A.
\newblock 2019.
\newblock {GANsynth}: Adversarial neural audio synthesis.
\newblock In {\em Proc. International Conference on Learning Representations}.

\bibitem[\protect\citeauthoryear{Griffin and Lim}{1984}]{griffin-lim}
Griffin, D., and Lim, J.
\newblock 1984.
\newblock Signal estimation from modified short-time fourier transform.
\newblock {\em IEEE Transactions on Acoustics, Speech, and Signal Processing}
  32(2):236--243.

\bibitem[\protect\citeauthoryear{Hawthorne \bgroup et al.\egroup
  }{2018}]{hawthorne18ismir}
Hawthorne, C.; Elsen, E.; Song, J.; Roberts, A.; Simon, I.; Raffel, C.; Engel,
  J.; Oore, S.; and Eck, D.
\newblock 2018.
\newblock Onsets and frames: Dual-objective piano transcription.
\newblock In {\em Proc. International Society for Music Information Retrieval
  Conference},  50--57.

\bibitem[\protect\citeauthoryear{Hono \bgroup et al.\egroup
  }{2018}]{sinsy18apsipa}
Hono, Y.; Murata, S.; Nakamura, K.; Hashimoto, K.; Oura, K.; Nankaku, Y.; and
  Tokuda, K.
\newblock 2018.
\newblock Recent development of the {DNN}-based singing voice synthesis
  system—{Sinsy}.
\newblock In {\em Proc. Asia-Pacific Signal and Information Processing
  Association Annual Summit and Conference},  1003--1009.

\bibitem[\protect\citeauthoryear{Hono \bgroup et al.\egroup }{2019}]{hono19}
Hono, Y.; Hashimoto, K.; Oura, K.; Nankaku, Y.; and Tokuda, K.
\newblock 2019.
\newblock Singing voice synthesis based on generative adversarial networks.
\newblock In {\em Proc. IEEE International Conference on Acoustics, Speech and
  Signal Processing},  6955--6959.

\bibitem[\protect\citeauthoryear{Hua and others}{2019}]{synthV}
Hua, K., et~al.
\newblock 2019.
\newblock {Synthesizer V}.
\newblock \url{https://synthesizerv.com}.

\bibitem[\protect\citeauthoryear{Imai}{1983}]{mlsa83icassp}
Imai, S.
\newblock 1983.
\newblock Cepstral analysis synthesis on the mel frequency scale.
\newblock In {\em Proc. IEEE International Conference on Acoustics, Speech and
  Signal Processing},  93--96.

\bibitem[\protect\citeauthoryear{Kalchbrenner \bgroup et al.\egroup
  }{2018}]{kalchbrenner18}
Kalchbrenner, N.; Elsen, E.; Simonyan, K.; Noury, S.; Casagrande, N.; Lockhart,
  E.; Stimberg, F.; van~den Oord, A.; Dieleman, S.; and Kavukcuoglu, K.
\newblock 2018.
\newblock {Efficient neural audio synthesis}.
\newblock In {\em Proc. International Conference on Machine Learning},
  2410--2419.

\bibitem[\protect\citeauthoryear{Kim \bgroup et al.\egroup }{2018a}]{kim18}
Kim, J.~W.; Salamon, J.; Li, P.; and Bello, J.~P.
\newblock 2018a.
\newblock {CREPE}: A convolutional representation for pitch estimation.
\newblock In {\em Proc. IEEE International Conference on Acoustics, Speech and
  Signal Processing},  161--165.
\newblock [Online] \url{https://github.com/marl/crepe}.

\bibitem[\protect\citeauthoryear{Kim \bgroup et al.\egroup
  }{2018b}]{kim18interspeech}
Kim, J.; Choi, H.; Park, J.; Hahn, M.; Kim, S.; and Kim, J.-J.
\newblock 2018b.
\newblock Korean singing voice synthesis system based on an {LSTM} recurrent
  neural network.
\newblock In {\em Proc. INTERSPEECH}.

\bibitem[\protect\citeauthoryear{Kingma and Ba}{2014}]{adam}
Kingma, D.~P., and Ba, J.
\newblock 2014.
\newblock Adam: A method for stochastic optimization.
\newblock {\em arXiv preprint arXiv:1412.6980}.

\bibitem[\protect\citeauthoryear{Kum and Nam}{2019}]{kum19as}
Kum, S., and Nam, J.
\newblock 2019.
\newblock Joint detection and classification of singing voice melody using
  convolutional recurrent neural networks.
\newblock {\em Applied Sciences} 9(7).
\newblock [Online] \url{https://github.com/keums/melodyExtraction_JDC}.

\bibitem[\protect\citeauthoryear{Lee \bgroup et al.\egroup
  }{2019}]{lee19interspeech}
Lee, J.; Choi, H.-S.; Jeon, C.-B.; Koo, J.; and Lee, K.
\newblock 2019.
\newblock Adversarially trained end-to-end {Korean} singing voice synthesis
  system.
\newblock In {\em Proc. INTERSPEECH}.

\bibitem[\protect\citeauthoryear{Lee, Choi, and Nam}{2018}]{lee18}
Lee, K.; Choi, K.; and Nam, J.
\newblock 2018.
\newblock Revisiting singing voice detection: a quantitative review and the
  future outlook.
\newblock In {\em Proc. International Society for Music Information Retrieval
  Conference},  506--513.

\bibitem[\protect\citeauthoryear{Liu and Yang}{2019}]{liu19}
Liu, J.-Y., and Yang, Y.-H.
\newblock 2019.
\newblock Dilated convolution with dilated {GRU} for music source separation.
\newblock In {\em Proc. International Joint Conference on Artificial
  Intelligence},  4718--4724.

\bibitem[\protect\citeauthoryear{Mao \bgroup et al.\egroup }{2017}]{lsgan}
Mao, X.; Li, Q.; Xie, H.; Lau, R.~Y.; Wang, Z.; and Smolley, S.~P.
\newblock 2017.
\newblock Least squares generative adversarial networks.
\newblock In {\em Proc. IEEE International Conference on Computer Vision}.

\bibitem[\protect\citeauthoryear{McFee \bgroup et al.\egroup
  }{2015}]{mcfee2015librosa}
McFee, B.; Raffel, C.; Liang, D.; Ellis, D. P. W.~.; McVicar, M.; Battenberg,
  E.; and Nieto, O.
\newblock 2015.
\newblock librosa: Audio and music signal analysis in {Python}.
\newblock In {\em Proc. Python in Science Conf.},  18--25.
\newblock [Online] \url{https://librosa.github.io/librosa/}.

\bibitem[\protect\citeauthoryear{Mirza and Osindero}{2014}]{cgan}
Mirza, M., and Osindero, S.
\newblock 2014.
\newblock Conditional generative adversarial nets.
\newblock {\em arXiv preprint arXiv:1411.1784}.

\bibitem[\protect\citeauthoryear{Nakamura \bgroup et al.\egroup
  }{2019}]{nakamura19arxiv}
Nakamura, K.; Hashimoto, K.; Oura, K.; and Yoshihiko~Nankaku, K.~T.
\newblock 2019.
\newblock Singing voice synthesis based on convolutional neural networks.
\newblock {\em arXiv preprint arXiv:1904.06868}.

\bibitem[\protect\citeauthoryear{Nishimura \bgroup et al.\egroup
  }{2016}]{nishimura16interspeech}
Nishimura, M.; Hashimoto, K.; Oura, K.; Nankaku, Y.; and Tokuda, K.
\newblock 2016.
\newblock Singing voice synthesis based on deep neural networks.
\newblock In {\em Proc. INTERSPEECH},  2478--2482.

\bibitem[\protect\citeauthoryear{Oura \bgroup et al.\egroup }{2010}]{ss_hmm2}
Oura, K.; Mase, A.; Yamada, T.; Muto, S.; Nankaku, Y.; and Tokuda, K.
\newblock 2010.
\newblock Recent development of the {HMM}-based singing voice synthesis
  system--sinsy.
\newblock In {\em Proc. ISCA Workshop on Speech Synthesis}.

\bibitem[\protect\citeauthoryear{Ozcan, Isikhan, and Alpkock}{2005}]{Ozcan2005}
Ozcan, G.; Isikhan, C.; and Alpkock, A.
\newblock 2005.
\newblock Melody extraction on {MIDI} music files.
\newblock In {\em Proc. IEEE International Symposium on Multimedia},  414--422.

\bibitem[\protect\citeauthoryear{Rafii \bgroup et al.\egroup }{2017}]{musdb18}
Rafii, Z.; Liutkus, A.; St{\"o}ter, F.-R.; Mimilakis, S.~I.; and Bittner, R.
\newblock 2017.
\newblock The {MUSDB18} corpus for music separation.

\bibitem[\protect\citeauthoryear{Saino \bgroup et al.\egroup }{2006}]{ss_hmm1}
Saino, K.; Zen, H.; Nankaku, Y.; Lee, A.; and Tokuda, K.
\newblock 2006.
\newblock An {HMM}-based singing voice synthesis system.
\newblock In {\em Proc. International Conference on Spoken Language
  Processing}.

\bibitem[\protect\citeauthoryear{Salimans and Kingma}{2016}]{salimans16}
Salimans, T., and Kingma, D.~P.
\newblock 2016.
\newblock {Weight normalization: A simple reparameterization to accelerate
  training of deep neural networks}.
\newblock {\em Proc. Advances in Neural Information Processing Systems}
  901--909.

\bibitem[\protect\citeauthoryear{van~den Oord \bgroup et al.\egroup
  }{2016}]{oord16}
van~den Oord, A.; Dieleman, S.; Zen, H.; Simonyan, K.; Vinyals, O.; Graves, A.;
  Kalchbrenner, N.; Senior, A.; and Kavukcuoglu, K.
\newblock 2016.
\newblock {WaveNet}: A generative model for raw audio.
\newblock {\em arXiv preprint arXiv:1609.03499}.

\bibitem[\protect\citeauthoryear{Vasquez and Lewis}{2019}]{melnet}
Vasquez, S., and Lewis, M.
\newblock 2019.
\newblock {MelNet}: A generative model for audio in the frequency domain.
\newblock {\em arXiv preprint arXiv:1906.01083}.

\bibitem[\protect\citeauthoryear{Wu and He}{2018}]{wu18}
Wu, Y., and He, K.
\newblock 2018.
\newblock Group normalization.
\newblock {\em arXiv preprint arXiv:1803.08494}.

\end{thebibliography}

\end{document}